\documentclass[iop]{emulateapj}
\usepackage{epsfig}
\usepackage{apjfonts}
\usepackage{amsmath,color,bm}
\usepackage{booktabs}
\usepackage[breaklinks,colorlinks,urlcolor=blue,citecolor=blue,linkcolor=blue]{hyperref}

\def\mc{\mathcal}

\def\Pr{{P}}
\def\vtheta{{\bm \theta}}
\def\H{{\mc{H}}}
\def\Hl{{\mc{H}_\textsc{l}}}
\def\Hu{{\mc{H}_\textsc{u}}}

\def\Blu{{\mc{B}_\textsc{u}^\textsc{l}}}
\def\Rlu{{\mc{R}_\textsc{u}^\textsc{l}}}

\def\Mchirp{{\mathcal{M}}}
\def\Msun{{M_{\odot}}}

\begin{document}

\title{Search for gravitational lensing signatures in LIGO-Virgo binary black hole events}

\author{O.A. Hannuksela$^1$, K. Haris$^2$, K.K.Y. Ng$^{3,4}$, S. Kumar$^{2,5,6}$, A.K. Mehta$^2$, D. Keitel$^7$, T.G.F. Li$^1$, P. Ajith$^{2,8}$}
\affiliation{$^1$~Department of Physics, Chinese University of Hong Kong, Sha Tin, Hong Kong}
\affiliation{$^2$~ International Centre for Theoretical Sciences, Tata Institute of Fundamental Research, Bangalore 560089, India}
\affiliation{$^3$~ LIGO, Massachusetts Institute of Technology, Cambridge, Massachusetts 02139, USA}
\affiliation{$^4$~Department of Physics and Kavli Institute for Astrophysics and Space Research, Massachusetts Institute of Technology, Cambridge, Massachusetts 02139, USA}
\affiliation{$^5$~ Max-Planck-Institut f\"ur Gravitationsphysik, Albert-Einstein-Institut, Callinstr. 38, 30167 Hannover, Germany}
\affiliation{$^6$~Leibniz Universit{\"a}t Hannover, 30167 Hannover, Germany}
\affiliation{$^7$~ University of Portsmouth, Institute of Cosmology and Gravitation, Portsmouth PO1 3FX, United Kingdom}
\affiliation{$^8$~Canadian Institute for Advanced Research, CIFAR Azrieli Global Scholar, MaRS Centre, West Tower, 661 University Ave, Toronto, ON M5G 1M1, Canada}

\begin{abstract}
We search for signatures of gravitational lensing in the binary black hole events detected by Advanced LIGO and Virgo during their first two observational runs. In particular, we look for three effects: 1) evidence of lensing magnification in the individual signals due to galaxy lenses, 2) evidence of multiple images due to strong lensing by galaxies, 3) evidence of wave optics effects due to point-mass lenses. We find no compelling evidence of any of these signatures in the observed gravitational wave signals. However, as the sensitivities of gravitational wave detectors improve in the future, detecting lensed events may become quite likely. 
\end{abstract}


\section{Introduction}
\label{sec:intro}
Advanced LIGO~\citep{TheLIGOScientific:2014jea,TheLIGOScientific:2016agk} and Virgo~\citep{TheVirgo:2014hva} have detected gravitational wave signals from ten binary black hole merger events during their first two observation runs, O1 and O2~\citep{LIGOScientific:2018mvr}. Upcoming observing runs will see further sensitivity upgrades to both LIGO and Virgo, as well as the prospects of a fourth detector, KAGRA~\citep{somiya:2012detector,aso2013interferometer,akutsu2018construction}, joining the network. A fifth detector is being built in India~\citep{M1100296}. As the sensitivities of these instruments improve, many novel avenues in astronomy research could become reality~\citep{abbott2018:prospects}. One such avenue is the study of gravitational lensing of gravitational waves. 

When gravitational waves propagate near massive astrophysical objects, their trajectories will curve, resulting in gravitational lensing~\citep{ohanian1974focusing,bliokh1975diffraction,bontz1981diffraction,thorne1983theory,deguchi1986diffraction,nakamura1998gravitational,takahashi2003gravitational}. Recent studies suggest that the resulting lensed gravitational waves could be detected by LIGO and Virgo as early as in the next few years~\citep{Ng:2017yiu, li2018gravitational}. Gravitational lensing, verified by numerous electromagnetic observations, has led to groundbreaking findings such as the detection of exoplanets~\citep{cassan2012one} and highly credible evidence for dark matter~\citep{clowe2004weak,markevitch2004direct}. Observation of lensed gravitational wave signals might present interesting applications in fundamental physics, astrophysics and cosmology; see, e.g,~\cite{Jung:2017flg, Lai:2018rto, dai2018detecting,2011MNRAS.415.2773S}.

Lensing could produce a number of observable effects on gravitational wave signals detectable by LIGO and Virgo. Firstly, a small fraction of binary black hole mergers will be strongly lensed by intervening galaxies~\citep{Ng:2017yiu}, and possibly by galaxy clusters~\citep{smith2018strong}. This would render detectable some of the binary black hole mergers that are beyond the horizon of Advanced LIGO and Virgo, due to the large lensing magnification~\citep{dai2017effect}. Since the mass scale of the lens is much larger than the gravitational wavelength, lensing does not affect the frequency profile of the signal in this case, which is referred to as the geometric optics limit. However, the overall magnification caused by lensing will be degenerate with the luminosity distance estimated from gravitational wave observations~\citep{Ng:2017yiu}. This will bias our estimation of the redshift to the binary and hence the intrinsic mass of the system. Thus, the lensed binaries would appear as a low redshift, high chirp mass population that could contradict known astrophysical binary mass models and, therefore, be potentially distinguishable as lensed events~\citep{dai2017effect, broadhurst2018reinterpreting}. Secondly, a fraction of the strongly lensed binary black hole merger events (by galaxy lenses) can produce multiple ``images'', which would arrive at the detector with relative time delays of minutes to weeks~\citep{2011MNRAS.415.2773S,Haris:2018vmn}. Thirdly, when the characteristic mass scale of the lens is comparable to the gravitational wavelength, interesting wave optics phenomena occur~\citep{nakamura1998gravitational,takahashi2003gravitational}. This can happen for the case of gravitational waves from stellar mass black hole mergers lensed by intermediate mass black holes \citep{Lai:2018rto}.  

We look for  evidence of the lensing effects mentioned above within the binary black hole events detected by Advanced LIGO/Virgo in the first and second observing run~\footnote{We do not include the binary neutron star merger event GW170817 in this study because the lensing probability is negligible at distances as small as 40 Mpc.}. We find that the LIGO/Virgo events are consistent with current astrophysical population models, and do not require lensing magnification to explain the observed mass and redshift distribution. Also, we find no conclusive evidence for multiple images by strong lensing nor the wave optics effects predicted in the limit of small lens masses. However, as the detector sensitivities improve, studying gravitational lensing with gravitational waves could soon become a realistic possibility. 

This paper is organized as follows: In Sec.~\ref{sec:magnification} we present results showing the lack of evidence of strong lensing magnification by modeling the high chirp mass, low redshift populations predicted by lensing and comparing them with LIGO/Virgo binary black hole measurements. In Sec.~\ref{sec:multipleimages} we use a Bayesian model selection to check if any pair of detected gravitational wave signals could be each others' strongly lensed counterpart. In Sec.~\ref{sec:waveoptics} we search for evidence of wave optics effects in observed signals by small point-like lenses using a templated search. We conclude and discuss future outlook in Sec.~\ref{sec:outlook}.

\section{No evidence of lensing magnification}
\label{sec:magnification}
In the regime of strong lensing by galaxies, lensing effects are well approximated by geometric optics. Due to the degeneracy between the distance and lensing magnification, it is difficult to distinguish whether a single binary black hole detection corresponds to an unlensed source at a distance $d_\mathrm{o}$ or a lensed image at $d_\mathrm{s} = d_\mathrm{obs}/\sqrt{\mu}$, where $\mu$ is the lensing magnification. 

Due to the cosmological expansion, the frequency of gravitational waves will be redshifted. Since gravitational wave frequencies are degenerate with the masses, what we estimate is the ``redshifted'' chirp mass $\mathcal{M}^z = (1+z) \, \mathcal{M}$, where $ \mathcal{M}$ is the intrinsic (true) chirp mass of the binary and $z$ the redshift. The estimated luminosity distance can be converted into a redshift estimate using a cosmological model, which can, in turn be used to estimate the intrinsic chirp mass $\mathcal{M}$ of the binary. The unknown lensing magnification will bias our estimation of the intrinsic mass and the distance (equivalently, the redshift) of the binary. Hence, lensed binaries will appear as a population of low redshift, high mass sources~\citep{dai2017effect}. \cite{broadhurst2018reinterpreting} argued that the high-mass events detected during the first observational run of LIGO are consistent with being strongly lensed. Here we demonstrate that the detections made during the first two observational runs of LIGO and Virgo do not show any statistically significant evidence of strong lensing.

We perform forward-modeling to predict the lensing rate of binary black holes observed by the Advance LIGO/Virgo detectors. Following~\cite{Ng:2017yiu}, the lensing rate $R_L$ is given by
\begin{align}
R_L=\int p_L(\mu, z_s) \, \frac{dN(\rho_L>\rho_\mathrm{th})}{dz_s \, d\vtheta} \, d\mu \, dz_s \, d\vtheta,
\end{align}
where $p_L(\mu, z_s)$ is the lensing probability at source redshift $z_s$ with magnification $\mu$, $\rho_L=\sqrt{\mu}\rho$ is the signal to noise ratio (SNR) of the lensed signal with $\rho$ as the SNR of the unlensed signal, $\rho_\text{th}$ is the network detection threshold and $\vtheta$ is the set of other binary parameters (component masses, spins, etc.). We set $\rho_{\text{th}}=10$ which is about the separation threshold between detections and marginal events in the GWTC-1 catalog~\citep{O2:CatalogUpdated}.

We simulate gravitational wave signals from an astrophysical distribution to forecast the rate of strongly lensed events. We distribute the binaries uniformly on the sky, with isotropic orientations, uniform spin magnitudes and isotropic spin directions. For the primary mass $m_1$, we use a power-law mass function $p(m_1)\propto m_1^{-2.35}$ with $5\Msun \leq m_1 \leq 50\Msun$. The power-law mass function follows from the initial mass function of progenitors~\citep{Salpeter}. The upper mass limit is motivated by pulsational pair-instability supernova, which prevents the formation of stellar remnants with mass $\sim 50-150 \Msun$~\citep{2017MNRAS.470.4739S,2017ApJ...836..244W,2016A&A...594A..97B,2002ApJ...567..532H}. The lower mass limit can be a consequence of rapid supernova mechanism, which explains the mass gap $\sim 2-5 \Msun$ in X-ray measurements~\citep{2012ApJ...757...91B,2012ApJ...749...91F,2010ApJ...725.1918O}. We determine the secondary mass $m_2$ by drawing from a uniformly distributed mass ratio $q=m_2/m_1$ with the constraints $m_1+m_2 \leq100\Msun$ and $5\Msun \leq m_2\leq m_1$. We use simulated redshift evolution from \citep{belczynski} as the merger rate density history. The redshift distribution is the product of merger rate density and differential comoving volume in Planck's $\Lambda\text{CDM}$ model~\citep{Planck15}. We then compute the optimal SNR $\rho$ of signals observed by the Advanced LIGO-Virgo network using the \textsc{IMRPhenomPv2} waveform family~\citep{Hannam:2013oca,IMRPhenomPv2,Husa:2015iqa}. We use publicly available noise power spectra of LIGO Hanford and LIGO Livingston in O1~\citep{LOSC} and O2~\citep{O2L1,O2H1}.

The magnification distribution of strong lensing is approximately $p(\mu) \propto \mu^{-3}$ for $\mu \geq 2$, at which $\mu = 2$ is the minimum magnification allowing multiple images \citep{lens_lectures}.We assume a constant comoving density of early-type galaxies as our lenses. We use singular isothermal sphere as our galactic lens model to determine the lensing probability at any magnification $\mu \geq 2$~\citep{turner},
\begin{align}
\tau(z_s | \mu \geq 2)=F\left( \frac{d_C(z_s)}{d_H} \right)^3,
\end{align}
where $d_H$ is the Hubble distance, $d_C$ is the comoving distance, and $F\sim0.0017$ is an empirical coefficient determined from galaxy surveys~\citep{Fcoeff,galaxylens}. Hence, the overall lensing probability is $p_L(\mu, z_s) = \tau(z_s | \mu \geq 2) \, p(\mu)$

We normalize the total rate (unlensed and lensed events) to our observation counts $\sim20$ per year, which is calculated from the coincidence analysis time in O1 and O2~\citep{catalog}. The resulting expected lensing rate is $\sim 0.1$ per year in O2 and $\sim 0.06$ per year in O1. We expect the low order-of-magnitude of lensing rates because the lensing optical depth $\tau(z_s)\sim\mathcal{O}(0.001)$ is a primary scaling factor to the lensing rate.

\begin{figure}
\centering
\includegraphics[width=\columnwidth]{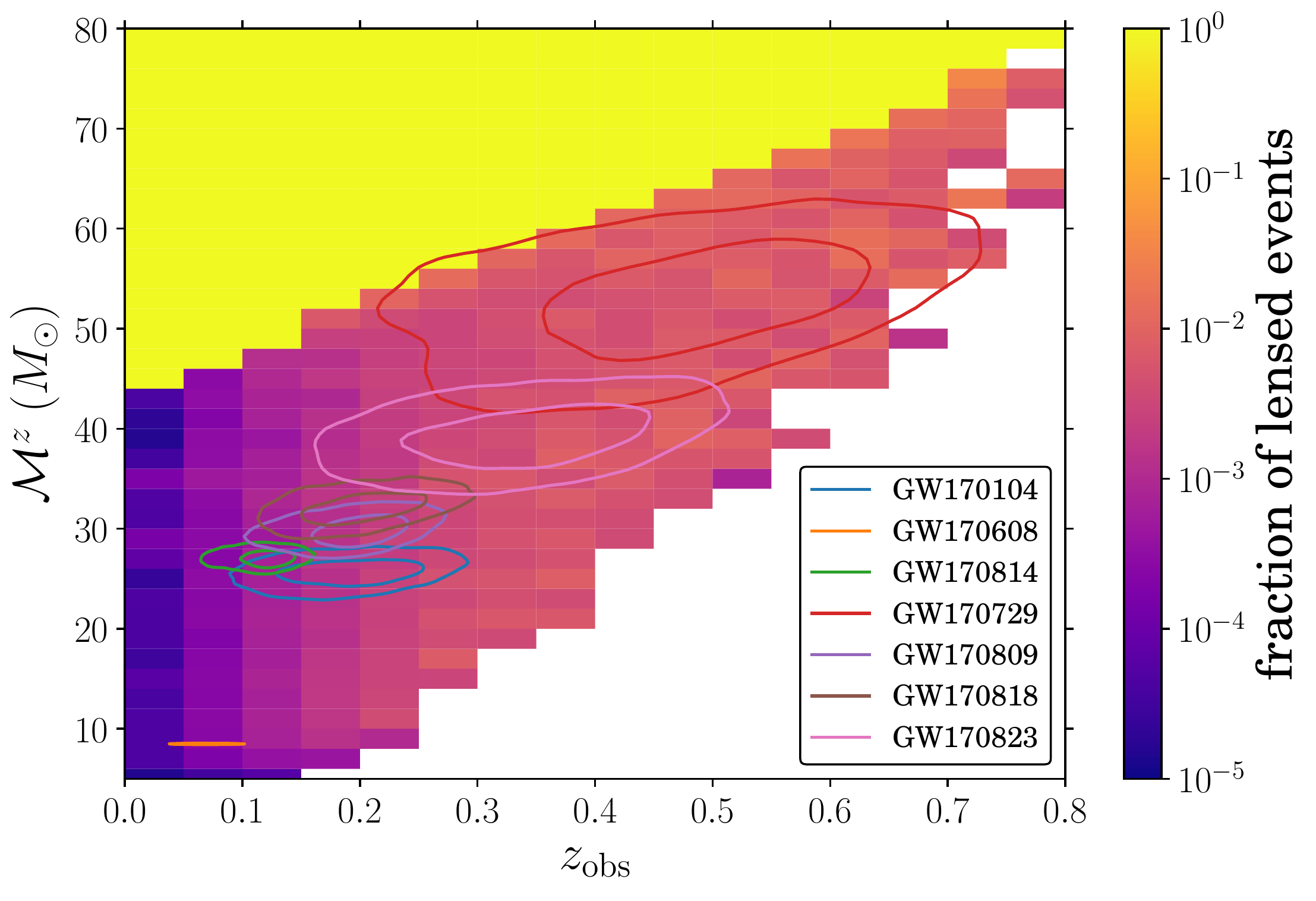}
\caption{Expected fraction of strongly lensed (magnified) over unlensed binary black hole mergers as a function of the observed redshift $z_\mathrm{obs}$ and redshifted chirp mass $\Mchirp^z$ in O2 sensitivity, obtained by forward modeling. The sharp transition from low fraction to unity at the high mass end is a consequence of the hard cut-off in intrinsic masses. The white region indicates no detection of lensed or unlensed events outside detector horizon. Contours of 50\% and 90\% confidence intervals of the posteriors of the binary black hole events from the first two observation runs of LIGO and Virgo are overlaid. The lensing probability is negligible ($\lesssim 10^{-2}$) in the region spanned by these posteriors, suggesting that these events are unlikely to be lensed.}
\label{mc-z}
\end{figure}

To further validate the low significance of strong lensing, we project the differential lensing rate on $\Mchirp^z-z_\mathrm{obs}$ plane, where $\Mchirp^z$ is the redshifted chirp mass and $z_\mathrm{obs}$ is the observed redshift after lensing, and calculate the fraction of lensing events in each bin (Fig.~\ref{mc-z}). All of the LIGO-Virgo detections lie inside the region of very low ($\lesssim 10^{-2}$) lensing probability.
From Fig.~\ref{mc-z}, the sharp transition of lensing fraction in high mass end implies that the upper mass limit is a significant indicator of lensing regime.
Even though we may observe events with masses falling in the lensing regime in Fig.~\ref{mc-z}, we emphasize that repeated binaries can also be an alternative explanation of high mass events exceeding upper mass limit~\citep{davide:2ndgen,2018PhRvL.120o1101R}.
This suggests that lensing is unnecessary to describe the population properties of these binary black hole events.

\section{No evidence of multiple images}
\label{sec:multipleimages}
A fraction of binary black hole mergers strongly lensed by galaxies would also be ``multiply imaged''~\citep{Ng:2017yiu}, with time delays of minutes to weeks between multiple images~\citep{Haris:2018vmn}. About $0.4\%$ of the black holes mergers are expected to produce detectable (SNR > 8) multiple images in Advanced LIGO-Virgo network at design sensitivity~\citep{Haris:2018vmn}. In this geometric optics regime, lensing only magnifies/demagnifies the lensed signals without affecting their frequency profile. Thus, posterior distributions of the intrinsic parameters that determine the frequency evolution of the signal (such as the redshifted masses and dimensionless spins of the black holes), estimated from multiple images, will be consistent with each other. Also, since the sky location of multiple images will be within the uncertainties of the gravitational wave sky localization, we can safely assume that the sky location estimated from multiple images will also be consistent. So will be the estimated inclination angle of the binary and the polarization angle. However, the estimated luminosity distance from the two images will in general be inconsistent since the distance is fully degenerate with the (unknown) magnification of the signal. 

From each pair of binary black hole signals detected by LIGO and Virgo, we compute the ratio of the marginalized likelihoods (Bayes factor) of the competing hypotheses: 1) that, the pair of signals are strongly lensed images of a single binary black hole merger, 2) that, they are produced by two independent mergers. This Bayes factor can be written as~\citep{Haris:2018vmn} 
\begin{equation}
\Blu = \int d \vtheta~ \frac{\Pr(\vtheta|d_1)~\Pr(\vtheta|d_2)}{\Pr(\vtheta)},
\label{eq:lensing_bayes_factor}
\end{equation}
where $\vtheta$ denotes the set of parameters that describes the signal (excluding the luminosity distance and arrival time), $\Pr(\vtheta)$ denotes the prior probability distribution of $\vtheta$, while $\Pr(\vtheta|d_{1})$ and $\Pr(\vtheta|d_{2})$ describe the posterior distributions of $\vtheta$ estimated from the data $d_1$ and $d_2$ containing the pair of signals under consideration. 

The measured time delay $\Delta t_0$ between two signals can also be used to compute the likelihood ratio of the two hypotheses. The Bayes factor between the lensed and unlensed hypotheses can be written as~\citep{Haris:2018vmn}
\begin{equation}
\Rlu = \frac{\Pr(\Delta t_0 |\Hl)}{\Pr(\Delta t_0|\Hu)}~,
\label{eq:bayesfactor_lensing_timedel}
\end{equation}
where $\Pr(\Delta t_0 |\H_A)$ with $A \in \{\textsc{l, u}\}$ is the prior distribution of $\Delta t$ (under the lensed or unlensed hypothesis) evaluated at $\Delta t = \Delta t_0$. The prior $\Pr(\Delta t_0 |\Hu)$ of the unlensed hypothesis is computed assuming that binary merger events follow a Poisson process. We use 714 days\footnote{ This is the total  duration from the beginning of O1 to the end of O2. In reality, the data is not available for the entire  714 days  due to the limited duty cycle of the Interferometers. We do not expect a significant change in  the prior distribution even if we include this correction.} as the  observation time for computing $\Pr(\Delta t_0 |\Hu)$. The prior distribution $\Pr(\Delta t|\Hl)$ of the time delay between strongly lensed signals is computed from an astrophysical simulation that employs reasonable distributions of galaxy lenses, mass function of binary black holes and redshift distribution of mergers, following~\cite{Haris:2018vmn}. 

We compute $\Blu$ from a pair of binary black hole signals by integrating the posterior distributions of the binary's parameters released by the LIGO-Virgo Collaboration~\citep{LIGOScientific:2018mvr,https://doi.org/10.7935/ksx7-qq51}. These posteriors are estimated by the \textsc{LALInferenceNest}~\citep{Veitch:2014wba,lalinference_o2} code using the gravitational waveform family \textsc{IMRPhenomPv2}. We use the joint posterior distributions of the following parameters $\vtheta := \{m_1^z, m_2^z, a_1, a_2, \cos \theta_{a1}, \cos \theta_{a2}, \alpha, \sin \delta, \theta_{J_N}\}$, where ${m_1^z, m_2^z}$ are the redshifted component masses, ${a_1, a_2}$ are the dimensionless spin magnitudes, ${\theta_{a1}, \theta_{a2}}$ are the polar angle of the spin orientations (with respect to the orbital angular momentum), ${\alpha, \sin \delta}$ denote the sky location, and $\theta_{J_N}$ is the orientation of the total angular momentum of the binary (with respect to the line of sight).~\footnote{\cite{Dai:2017huk} have discovered that, if we neglect the effects of spin precession and non-quadrupole modes, multiple images are related to each other by specific phase shifts. Hence the consistency of the coalescence phase $\phi_c$ and polarization angle $\psi$, which is degenerate with $\phi_c$ can also be used to determine the consistency of multiple images. However, we are using a more general waveform family that include spin precession, where such a relationship does not hold. Hence we do not check the consistency of $\phi_0$ and $\psi$.} The Bayes factor in Eq.(\ref{eq:lensing_bayes_factor}) is computed by numerically integrating the products of the Gaussian kernel density estimates of the posterior distributions of $\vtheta$ from each pair of events, after marginalizing them over all other parameters using standard priors in the LIGO-Virgo parameter estimation~\citep{LIGOScientific:2018mvr}. 

\begin{figure}
\includegraphics[width=\linewidth]{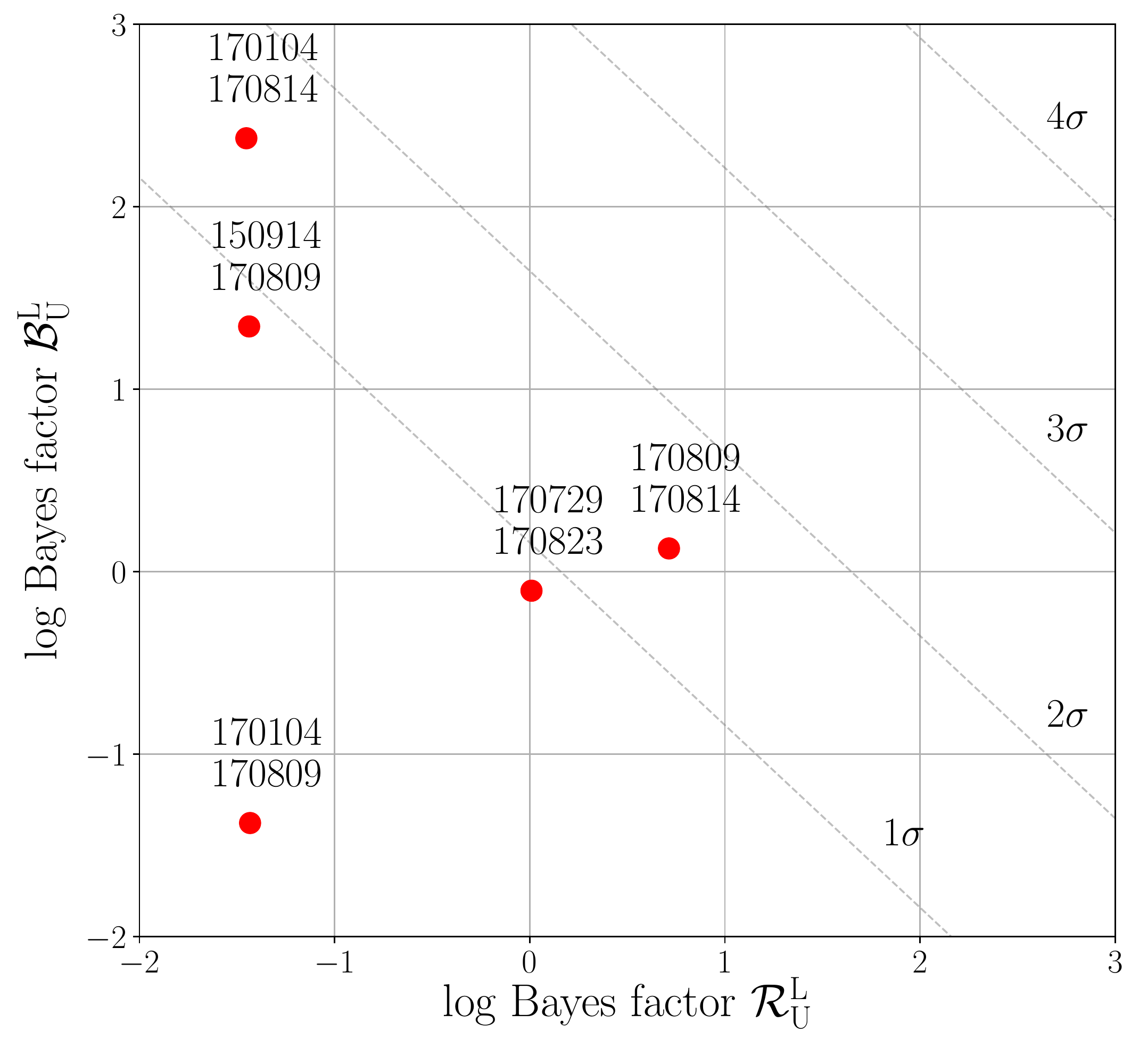}
\caption{Scatter plot of the $\mathrm{log}_{10}$ Bayes factors $\Blu$ computed from the consistency of posteriors of signal parameters estimated from each pair of binary black holes events and Bayes factors $\Rlu$ computed from the time delay between pairs of events. The significance of these Bayes factors is shown by dashed lines (in terms of Gaussian standard deviations). This is estimated by performing simulations of unlensed events in simulated Gaussian noise and estimating the probability of unlensed events producing Bayes factors of this value.  In summary, we do not see any strong evidence for multiply lensed images in LIGO-Virgo binary black hole detections. Note that, out of 45 event pairs, only those pairs with $\mathrm{log}_{10}$ Bayes factors greater than $-2$ are shown in the plot. We have taken into account the effect of trials factor due the 45 event pairs.}
\label{fig:bayesfactor_multiimage}
\end{figure}
Figure~\ref{fig:bayesfactor_multiimage} presents a scatter plot of the Bayes factors $\Blu$ and $\Rlu$ computed from binary black hole event pairs observed by LIGO and Virgo during the first two observation runs. 
Since the $\Blu$ and $\Rlu$ are computed using unrelated information, we can compute a joint Bayes factor by multiplying $\Blu$ and $\Rlu$, which is used to determine the significance for each pair~\citep{Haris:2018vmn}.
Figure~\ref{fig:bayesfactor_multiimage} also shows the significance of these Bayes factor values, $\Blu \times \Rlu$, in terms of Gaussian standard deviations. The significance is estimated from simulations of unlensed binary black hole events in Gaussian noise with power spectra of the Advanced LIGO-Virgo network with design sensitivity, presented in~\cite{Haris:2018vmn}~\footnote{The significance of lensed event pairs will be even lower if we used the actual O1-O2 noise spectra, due to the lower sensitivity. Hence this is an optimistic estimate of the significance of these Bayes factors. Note that we have also taken into account the trials factor from the 45 event pairs, from 10 events.}. 

The event pairs GW170104-GW170814 and GW150914-GW170809 show the highest Bayes factors $\Blu \sim {198}$ and ${29}$ --- their posteriors overlap at a reasonable confidence level to suggest a possible explanation of them as double images of a single source based on waveform similarity (see Figs.~\ref{fig:corner_170104_170814} and \ref{fig:corner_150914_170809} of the Appendix). However, galaxy lenses are unlikely to produce time delays as long as 7 or 23 months between the images~\citep{Haris:2018vmn}, resulting in a small $\Rlu\sim 4 \times 10^{-3}$ and $10^{-4}$ for both pairs. If galaxy clusters were a viable lensing source, then one could expect time delays of a few months~\citep{Smith:2017mqu,smith2018strong}. However,  the rate of strongly lensed binary black hole mergers  by galaxy clusters at current sensitivity is around $10^{-5}$ per year \citep{smith2018deep}, disfavoring this scenario. On the other hand, the time delay between GW170809 and GW170814 is consistent with galaxy lenses ($\Rlu \sim {3.3}$). While the projected 1-dimensional posterior of, e.g., chirp mass overlap within 90\% confidence~\citep{Broadhurst:2019ijv}, this is mainly caused by correlation with other intrinsic parameters, e.g. effective spin. The posteriors in higher dimensions do not show similar overlap (see Fig.~\ref{fig:corner_170809_170814} of the Appendix), implying that these waveforms can be discriminated from each other with reasonable confidence. Indeed, a full higher-dimensional consistency check between the estimated parameters from this pair does not significantly favor lensing  ($\Blu \sim {1.2}$). The joint Bayes factors $\Blu \times \Rlu$ for these pairs are $0.9$ (GW170104-GW170814), $4 \times 10^{-3}$ (GW150914-GW170809) and $4$ (GW170809-GW170814). In summary, we do not see any strong evidence for the hypothesis that any of the pairs of binary black hole signals are lensed images of the same merger event. We have also repeated the same calculation employing the waveform family \textsc{SEOBNRv3}~\citep{Pan:2013rra,Taracchini:2013rva,Babak:2016tgq}. The Bayes factors that we obtain from this analysis are consistent with those presented in Fig.~\ref{fig:bayesfactor_multiimage}.

We also compute the Bayes factor of the hypothesis that there exists at least one multiply imaged event in the entire catalog of events observed by Advanced LIGO-Virgo in the first and second observing run (without specifically identifying that pair). Considering the fact that the probability for observing more than 2 lensed images of a single merger is negligible, the joint Bayes factor
$ \sum_{p\in\mathrm{pairs}}\Blu(p) \Rlu(p) $
is equal to $5.2$, and is not highly significant.

\section{No evidence of wave optics effects}
\label{sec:waveoptics}
When a gravitational wave propagates around an object of size similar to its wavelength, interesting wave optics effects are produced due to the superposition of several lensed wavefronts with variable magnifications and time-delays~\citep{ohanian1974focusing,bliokh1975diffraction,bontz1981diffraction,thorne1983theory,deguchi1986diffraction,nakamura1998gravitational,takahashi2003gravitational,lensingstellar2018}. In such a scenario, the observed waveform will have characteristic beating patterns detectable in LIGO and Virgo~\citep{cao2014gravitational,Lai:2018rto}, if the lensing object's mass $M_L \lesssim 10^5 M_\odot$, e.g., that of intermediate-mass black holes. Such lensing effects could be detected if the lens lies close to a caustic and its effective Einstein radius is expanded~\cite[see][for more details]{Lai:2018rto}. We search for such lensing effects in the LIGO-Virgo detections, assuming point-like lenses as considered in~\cite{Lai:2018rto}. 

The effect of lensing may be solved from the Einstein field equations, when the gravitational potential is too weak to change the polarization of the wave ($U\ll 1$), and when the gravitational wave can be separated from the background space-time~\citep{nakamura1998gravitational,takahashi2003gravitational}~\footnote{When the wavelength of the gravitational wave is much larger than, the object's size and the wave travels near the object, the wave may no longer be separated from the background and wave scattering occurs (see e.g.~\cite{takahashi2005scattering}). We do not consider this effect.}. Such an approximation is valid when the lensing object's size is comparable to, or larger than the wavelength of the gravitational wave. The result in the point mass lens approximation yields a frequency dependent {magnification factor} $F(f; M_{L}^z,y)$, that is a function of the redshifted lens mass $M_{L}^z$ and source position $y=D_L \eta /\xi_0 D_S$, where $D_L$ and $D_S$ are angular diameter distances of the lens and the gravitational wave source, respectively, $\eta$ is the distance to the source from the line-of-sight of the lens and $\xi_0$ is the lens' Einstein radius~\citep{nakamura1998gravitational,takahashi2003gravitational,Lai:2018rto}. The magnification factor transforms an unlensed waveform to a lensed waveform. 

We search for signatures of point mass lenses within a range of source positions $y\in [0.1,3]$ and redshifted mass of the lens $M_{L}^z \in [1,M_{\max}] \rm M_\odot$ in all O1 and O2 detections using \textsc{LALInferenceNest} and lensed \textsc{IMRPhenomPv2} waveform family, as implemented as \cite{Lai:2018rto}. Our upper bound for the lens mass, $M_{\rm max}$, is chosen so that the time-delay between the two lensed images is large enough for the lensed waves to be well-separated~\citep{takahashi2003wave}, and we assume agnostically that the lens can be in any mass range and hence choose a uniform prior in $\log_{10}~M_{L}^z$. Furthermore, we cut off the source position $y$ at $3$, because the lensing effects beyond this point are unmeasurable, while at $y\lesssim 0.1$ the lensing probability is small. The prior $p(y) \propto y$ is chosen based on geometrical argument and isotropy, i.e., the probability distribution for the line-of-sight distance goes as $p(<\eta) \propto \pi \eta^2$, and we have verified that this prior is largely unaffected by the assumption for the astrophysical distribution of lenses. For additional details of the lensing formalism and the choice of prior, refer to~\cite{Lai:2018rto}. 

\begin{figure}
 \centering 
 \includegraphics[width=\columnwidth]{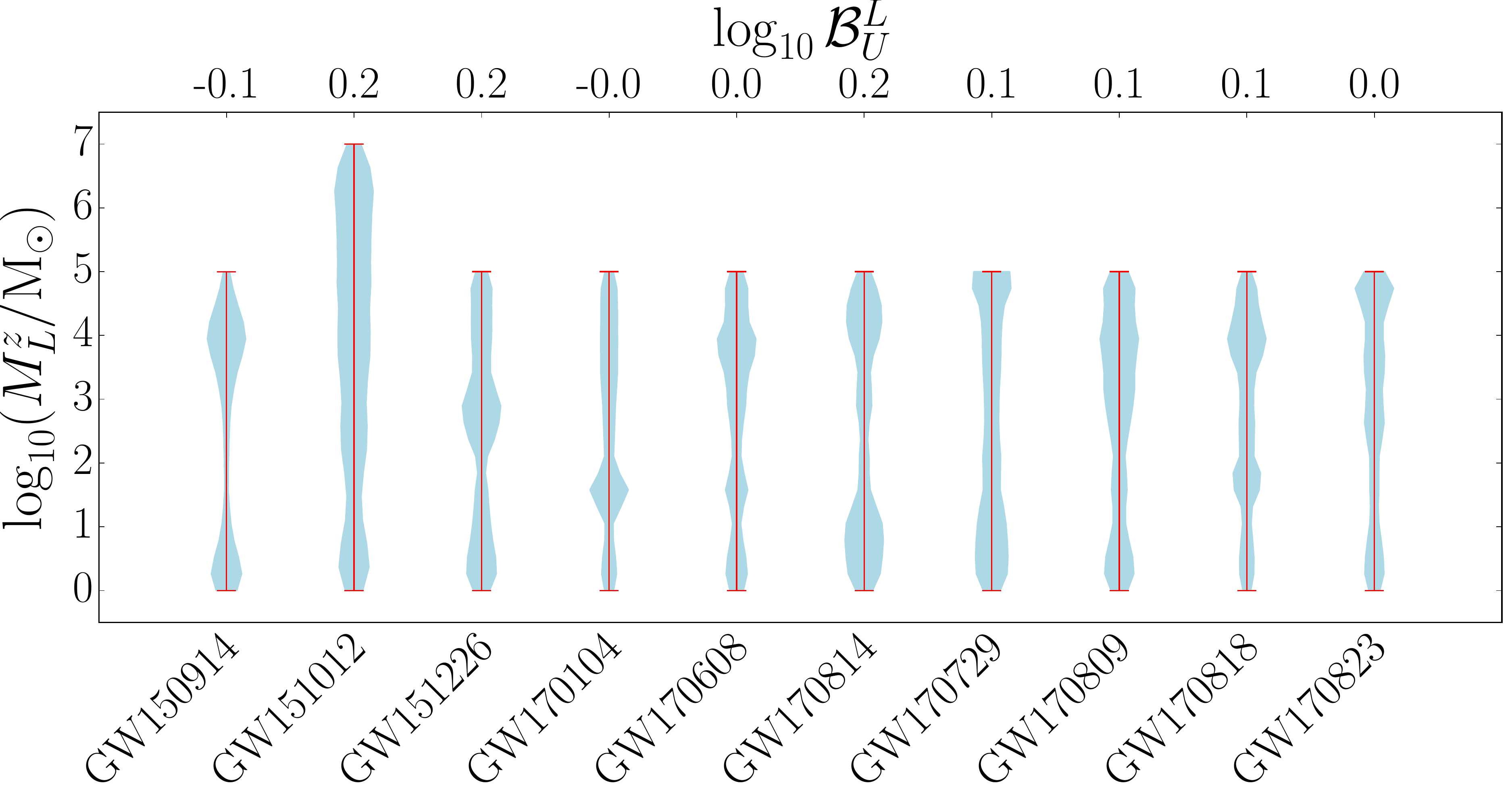}
 \caption{Posterior distribution of redshifted lens mass $M_L^z$ (violin plots) and the log Bayes factor between lensed and unlensed hypothesis $\log_{10} \mathcal{B}^L_U$ (top vertical axis) for wave optics effects in each gravitational wave event. The Bayes factors and the lens mass posteriors have been computed using nested sampling assuming a $\log$-uniform redshifted lens mass prior. None of the Bayes factors are significant enough to favor the lensing hypothesis.}
 \label{fig:waveopticsbayesianfactors}
\end{figure}

We compute the ratio of the Bayesian evidences of the lensed and unlensed hypotheses (using lensed and unlensed waveforms, respectively) estimated using \textsc{LALInferenceNest}. Figure~\ref{fig:waveopticsbayesianfactors} shows the marginalized posterior distributions of redshifted lens mass $M_L^z$ (violin plots) and the Bayes factors between the lensed and unlensed hypothesis $\mathcal{B}^{L}_{U}$ for each gravitational wave event (note that the precise definition of this Bayes factor is different from that of Sec.~\ref{sec:multipleimages}, although we use the same notation). The posterior distributions do not peak at zero lens mass due to the free source position variable $y$, which at higher values reduce the lensing effect, causing the lens mass posterior to be broad instead. Note that for the GW151012 event we have made the prior broader as the peak of the posterior was otherwise not captured. We find that the Bayes factor $\log_{10} \mathcal{B}^{L}_{U} < 0.2$ for all events. Hence, we find no evidence to support the lensing hypothesis by smaller point-like lenses.

\section{Outlook}
\label{sec:outlook}
We have searched for lensing effects in the binary black hole observations by LIGO and Virgo during the observing runs O1 and O2, finding no strong evidence of gravitational lensing. In particular, we looked for three effects. Firstly, we searched for evidence of high lensing magnification in the observed signals by comparing the chirp mass --- red shift distribution of observed binary black holes to the statistically predicted populations of lensed and unlensed signals. Secondly, we looked for evidence of multiply imaged signals by investigating the consistency of the estimated parameters among all pairs of events. Thirdly, we looked for evidence of wave optics effects in the observed signals by point-like lenses. None of these investigations revealed any lensing effects in the observed signals. 

While the probability of lensed gravitational waves is low, in the future, as detector sensitivities improve further, it will become increasingly possible to observe strong lensing~\citep{Ng:2017yiu}. Since Advanced LIGO and Virgo are expected to observe hundreds of binary black hole mergers as they reach their design sensitivity, according to current estimates, more than one strongly lensed signal will be observable per year. Apart from verifying a fundamental prediction of general relativity using a messenger that is different from electromagnetic waves, such an observation might enable precision localization of the merger when combined with optical observations of the lens galaxy~\citep{Mehta:2019aa}. Since the fraction of lensed events will be small, we do not expect lensing to introduce significant biases in population analysis.

Detecting wave optics effects, e.g. by intermediate-mass black holes, could be possible at least in the future third generation detectors~\citep{Lai:2018rto,lensingstellar2018}, but detection rates are highly uncertain in the current ground-based detectors. However, it is worth noting that the time-resolution of LIGO would be able to probe lensing that are below the typical angular resolution of optical or radio telescopes, and hence could uncover hidden lens populations that could have been missed. The prime targets for weak lensing are likely to be smaller substructures that would be enhanced by the galaxies' potential, which have been observed in the optical band~\citep{diego2018dark}. Indeed, lensing observations of gravitational waves are likely to become a powerful tool for astronomy in the coming years.

\bigskip 
\paragraph{Acknowledgments:} We thank the LIGO Scientific Collaboration and Virgo Collaboration for providing the data of binary black hole observations during the first two observation runs of Advanced LIGO and Virgo. PA's research was supported by the Science and Engineering Research Board, India through a Ramanujan Fellowship, by the Max Planck Society through a Max Planck Partner Group at ICTS-TIFR, and by the Canadian Institute for Advanced Research through the CIFAR Azrieli Global Scholars program. SK acknowledges support from national post doctoral fellowship (PDF/2016/001294) by Scientific and Engineering Research Board, India. OAH is supported by the Hong Kong Ph.D. Fellowship Scheme (HKPFS) issued by the Research Grants Council (RGC) of Hong Kong. The work described in this paper was partially supported by grants from the Research Grants Council of the Hong Kong (Project No. CUHK 14310816, CUHK 24304317 and CUHK 14306218) and the Direct Grant for Research from the Research Committee of the Chinese University of Hong Kong. KKYN acknowledges support of the National Science Foundation, and the LIGO Laboratory. LIGO was constructed by the California Institute of Technology and Massachusetts Institute of Technology with funding from the National Science Foundation and operates under cooperative agreement PHY-0757058. Computations were performed at the ICTS cluster Alice and the LIGO Hanford cluster. KH, SK, AKM and PA thank Tejaswi Venumadhav,  B. Sathyaprakash, Jolien Creighton, Xiaoshu Liu, Ignacio Magana Hernandez and Chad Hanna for useful discussions. OAH, KKYN and TGFL also acknowledge useful input from Peter~T.~H.~Pang, Rico K.~L.~Lo.

\bibliographystyle{apj}
\bibliography{cbc-group,lensing,waveoptics}

\newpage
\bigskip 
\appendix 
\section{Detailed investigations of event pairs showing marginal evidence of lensing}
\label{sec:appendix}

\begin{figure*}[tbh]
\includegraphics[width=0.8\linewidth]{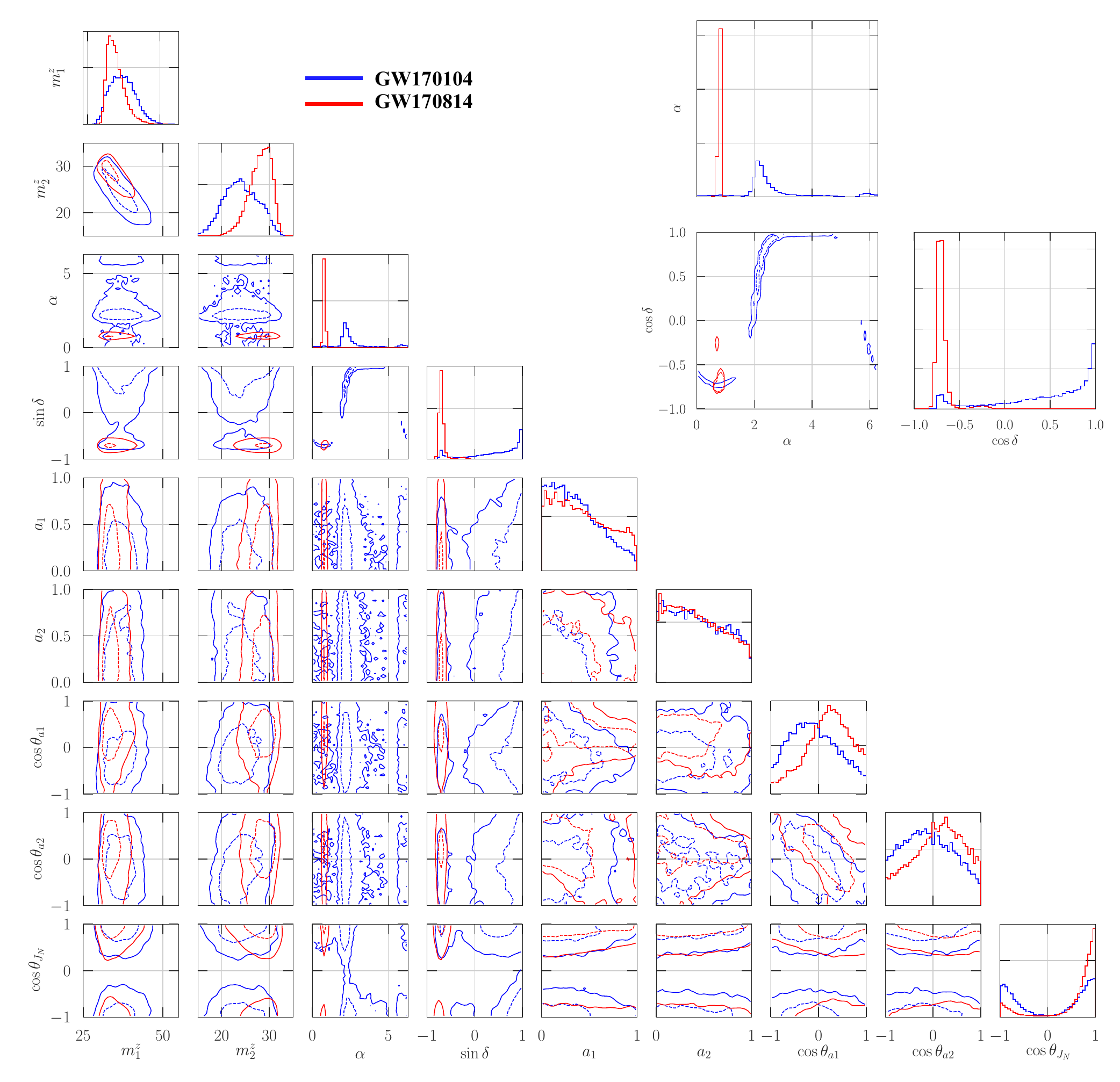}
\caption{Marginalized 2D and 1D posterior distributions of the parameters that are included in the consistency test, for the event pair GW170104 (blue), GW170814(red). Here, ${m_1^z, m_2^z}$ are the redshifted component masses, ${a_1, a_2}$ are the dimensionless spin magnitudes, ${\theta_{a1}, \theta_{a2}}$ are the polar angle of the spin orientations (with respect to the orbital angular momentum), ${\alpha, \sin \delta}$ denote the sky location, and $\theta_{J_N}$ is the orientation of the total angular momentum of the binary (with respect to the line of sight).  The solid (dashed) condors corrsponds to the $90\%(50\%)$ confidence levels of the 2D distributions. The inset plot shows  the marginalized posterior distributions of the sky localization parameters for these events. Overall, the posteriors have some levels of overlap, thus resulting in a considerable Bayes factor of $\Blu \sim {198}$  supporting the lensing hypothesis, purely based on parameter consistency. However, galaxy lenses are unlikely to produce time delay of 7 months between the images, resulting in a small Bayes factor $\Rlu\sim 4 \times 10^{-3}$ based on time delay considerations.}
\label{fig:corner_170104_170814}
\end{figure*}

Here we present additional investigations on the event pairs that show marginal evidence of multiply-imaged lensing in the analysis presented in Sec.~\ref{sec:multipleimages}, providing a qualitative explanation of the Bayes factors presented in that section in terms of the overlap of the estimated posteriors from these event pairs. Figure~\ref{fig:corner_170104_170814} presents the 2D and 1D marginalized posterior distributions of the parameters that are included in the consistency test, for the event pair GW17014-GW170814. Posteriors have appreciable levels of overlap in many parameters, thus resulting in a considerable Bayes factor of $\Blu \sim {198}$ supporting the lensing hypothesis, purely based on parameter consistency. However, galaxy lenses are unlikely to produce time delay of 7 months between the images~\citep{Haris:2018vmn}, resulting in a small Bayes factor $\Rlu\sim 4 \times 10^{-3}$ based on time delay considerations. 

\begin{figure*}[tbh]
\includegraphics[width=0.8\linewidth]{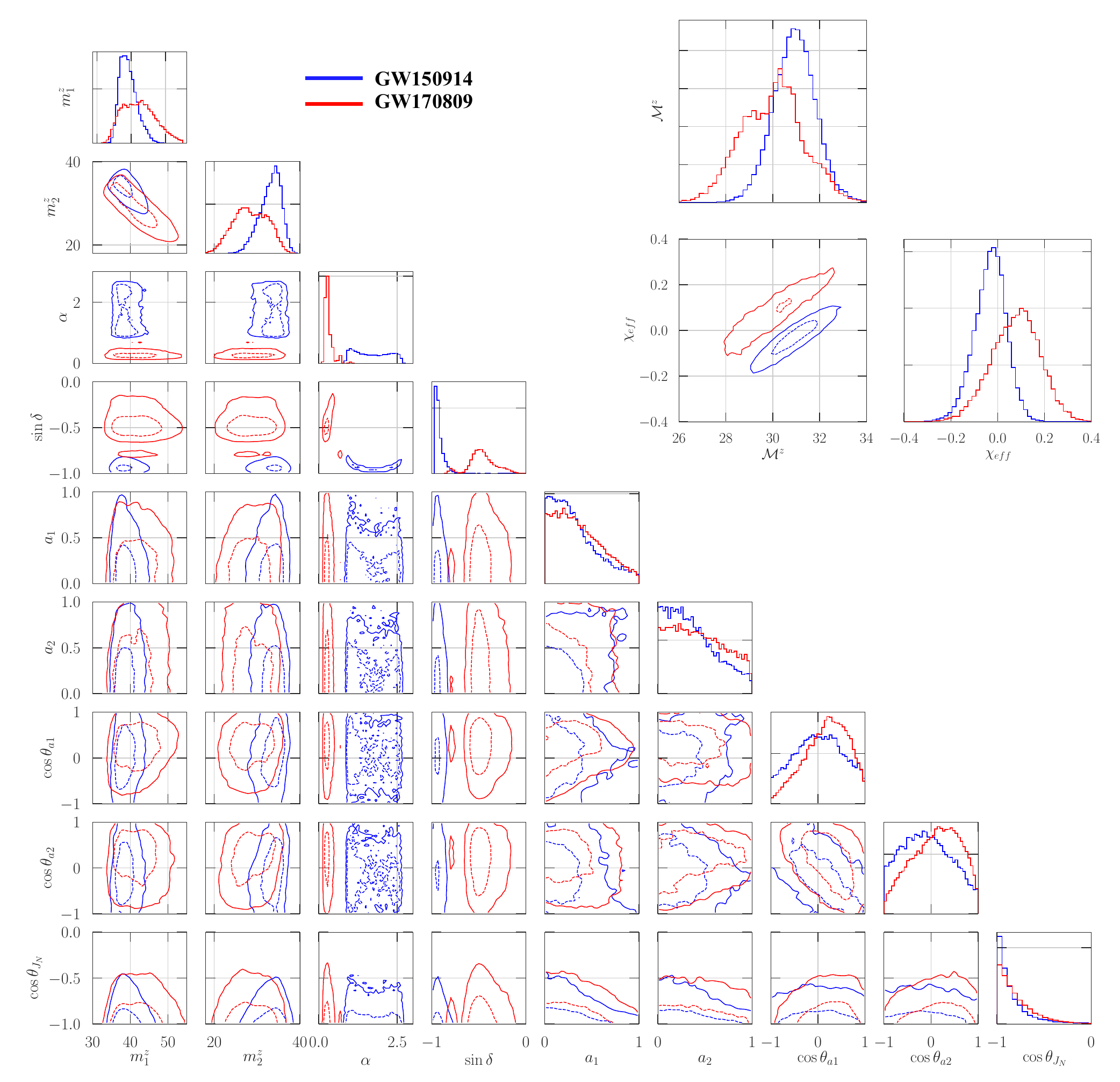}
\caption{Same as Fig.~\ref{fig:corner_170104_170814}, except that the figure corresponds to the 150914 (blue), GW170809 (red) event pair. The inset plot shows  the marginalized posterior distributions of the  redshifted chirp mass $\mathcal{M}^z$ and effective spin $\chi_{\rm eff}$ for these events. Marginalized 1D posteriors have some levels of overlap in many parameters; however 2D posteriors show good separation in many parameters, e.g., in $\mathcal{M}^z - \chi_{\rm eff}$. The resulting Bayes factor supporting the lensing hypothesis, based on parameter consistency is $\Blu \sim {29}$. However, galaxy lenses are unlikely to produce time delay of 23 months between the images, resulting in a small Bayes factor $\Rlu \sim \times 10^{-4}$ based on time delay considerations.}
\label{fig:corner_150914_170809}
\end{figure*}

\begin{figure*}[tbh]
\includegraphics[width=0.8\linewidth]{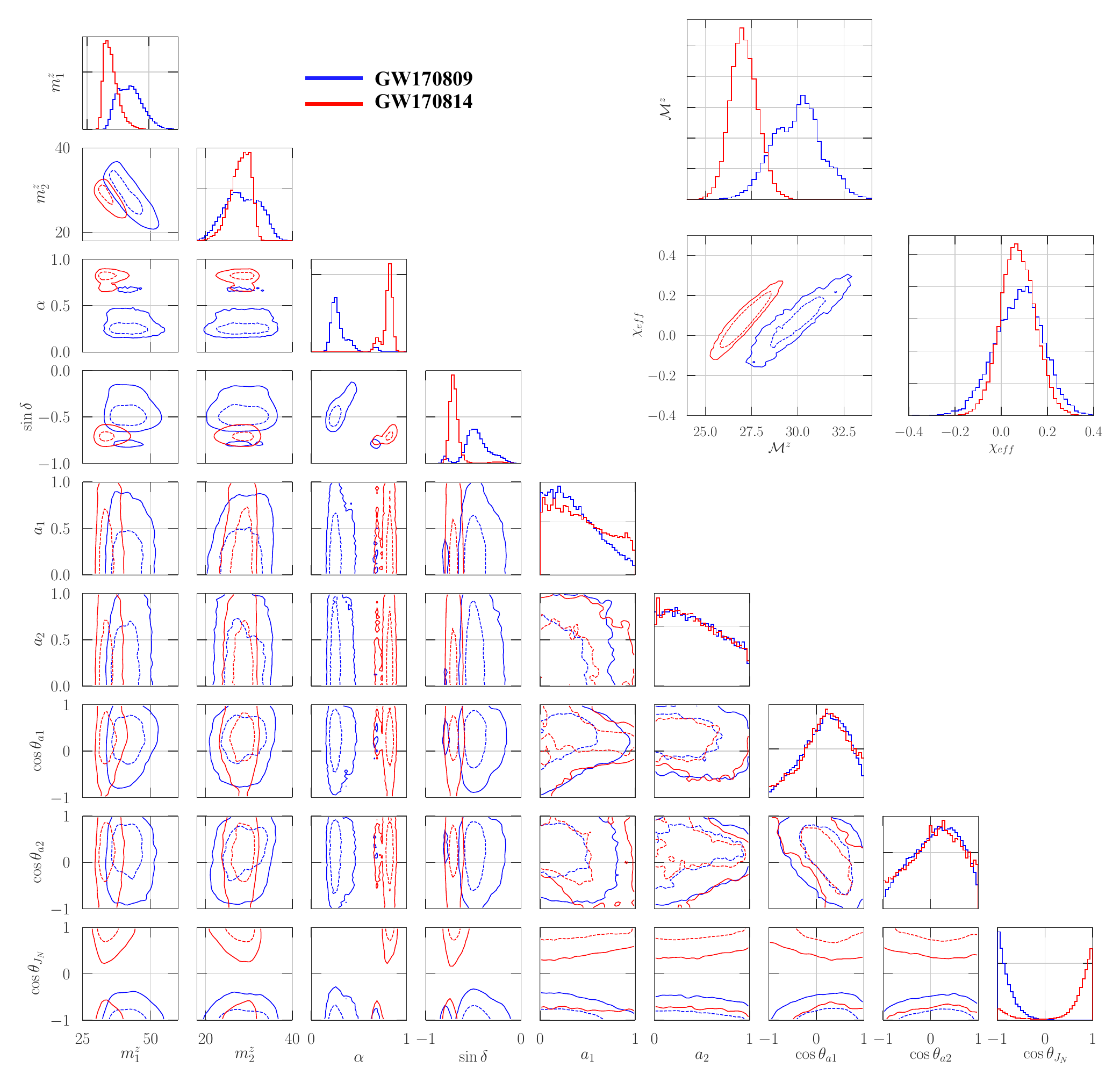}
\caption{Same as Fig.~\ref{fig:corner_170104_170814}, except that the figure corresponds to the GW170809 (blue), GW170814 (red) event pair. Marginalized 1D posteriors have some levels of overlap in many parameters; however 2D posteriors show good separation in many parameters, e.g., in $\mathcal{M}^z - \chi_{\rm eff}$. The resulting Bayes factor supporting the lensing hypothesis, based on parameter consistency is $\Blu \sim {1.2}$.}
\label{fig:corner_170809_170814}
\end{figure*}

Figure~\ref{fig:corner_150914_170809} shows similar plots for the event pair GW150914-GW170809. Although marginalized 1D posteriors have some levels of overlap in many parameters, 2D posteriors show good separation in many parameters, e.g., in $\mathcal{M}^z - \chi_{\rm eff}$. The resulting Bayes factor supporting the lensing hypothesis, based on parameter consistency is $\Blu \sim {29}$. However, galaxy lenses are unlikely to produce time delay of 23 months between the images, resulting in a small Bayes factor $\Rlu \sim \times 10^{-4}$ based on time delay considerations. Figure~\ref{fig:corner_170809_170814} shows similar plots for the event pair GW170809-GW170814. Here also, the 2D posteriors of several parameters (e.g., in $\mathcal{M}^z - \chi_{\rm eff}$) show poor overlaps, suggesting that the full multidimensional posteriors do not have significant overlap. The resultant Bayes factor for parameter consistency is $\Blu \sim {1.2}$, even though, the time delay between these events is consistent with galaxy lenses, producing a Bayes factor $\Rlu \sim {3.3}$ based on time delay.

\bigskip

\end{document}